# Chapter 1

# Big Data Analytics = Machine Learning + Cloud Computing

**Caesar Wu, Rajkumar Buyya, and Kotagiri Ramamohanarao**

**Abstract:** Big Data can mean different things to different people. The scale and challenges of Big Data are often described using three attributes, namely Volume, Velocity and Variety (3Vs), which only reflect some of the aspects of data. In this chapter we review historical aspects of the term "Big Data" and the associated analytics. We augment 3Vs with additional attributes of Big Data to make it more comprehensive and relevant. We show that Big Data is not just 3Vs, but $3^2$ Vs, that is, 9 Vs covering the fundamental motivation behind Big Data, which is to incorporate Business Intelligence (BI) based on different hypothesis or statistical models so that Big Data Analytics (BDA) can enable decision makers to make useful predictions for some crucial decisions or researching results. History of Big Data has demonstrated that the most cost effective way of performing BDA is to employ Machine Learning (ML) on the Cloud Computing (CC) based infrastructure or simply, ML + CC → BDA. This chapter is devoted to help decision makers by defining BDA as a solution and opportunity to address their business needs.

## 1.1    Introduction

Although the term Big Data has become popular, there is no general consensus about what it really means. Often, many professional data analysts would imply the process of Extraction, Transformation and Load (ETL) for large datasets as the connotation of Big Data. A popular description of Big Data is based on three attributes of data: volume, velocity, and variety (or 3Vs). Nevertheless, it does not capture all the aspects of Big Data accurately. In order to provide a comprehensive meaning of Big Data, we will investigate this term from a historical perspective and see how it has been evolving from yesterday's meaning to today's connotation.

Historically, the term Big Data is quite vague and ill-defined. It is not a precise term and does not carry a particular meaning rather than the notion of its size. The word "Big" is too generic. The question how "big" is big and how "small" is small [1] is relative to time, space and a circumstance. From an evolutionary perspective, the size of "Big Data" is always evolving. If we use the current global Internet traffic capacity [2] as a measuring stick yard, the meaning of Big Data's volume would lie between Terabyte (TB or $10^{12}$ or $2^{40}$) and Zettabyte (ZB or $10^{21}$ or $2^{70}$) range. Based on historical data traffic growth rate, Cisco claimed that human has entered the ZB era in 2015 [2]. To understand significance of the data volume's impact, let us glance at the average size of different data files shown in Table 1.

| Media | Average Size of Data File | Notes (2014) |
|---|---|---|
| Web Page | 1.6 - 2 MB | Ave 100 objects |
| eBook | 1 - 5 MB | 200-350 pages |
| Song | 3.5 - 5.8 MB | Ave 1.9 MB/per minute(MP3) 256 Kbps rate (3 mins) |
| Movie | 100 - 120 GB | 60 frames per second (MPEG-4 format, Full High Definition, 2 hours) |

Table 1: Typical Size of Different Data Files

The main aim of this chapter is to provide a historical view of Big Data and to argue that Big Data is not just 3Vs, but rather $3^2$Vs or 9Vs. These additional Big Data attributes reflect the real motivation behind Big Data Analytics (BDA). We believe that these expanded features clarify some basic questions about the essence of BDA: what problems Big Data can address, and what problems should not be confused as BDA. These issues are covered in the chapter through analysis of historical developments along with associated technologies that support Big Data processing. The rest of the chapter is organised into eight sections as follows:

1) A historical Review for Big Data
2) Interpretation of Big Data 3Vs, 4Vs and 6Vs
3) Defining Big Data from 3Vs to 3²Vs
4) Big Data and Machine Learning
5) Big Data and Cloud Computing
6) Hadoop, HDFS, MapReduce, Spark and Flink
7) ML + CC → BDA and Guidelines
8) Conclusion

## 1.2    A Historical Review of Big Data

In order to capture the essence of Big Data, we provide the origin and history of BDA and then propose a precise definition of BDA.

### 1.2.1    The Origin of Big Data

Several studies have been conducted on historical views and developments in BDA area.  Gil Press [3] provided a short history of Big Data starting from 1944, which was based on Rider's work [4]. He covered 68 years of history of evolution of Big Data between 1944 and 2012 and illustrated 32 Big Data related events in the recent data science history. As Press' indicated in his article, the fine line between the growth of data and Big Data has become blurred. Very often, the growth rate of data has been referred as "information explosion" (although "data" and "information" are often used interchangeably, two terms have different connotations). Press' study is quite comprehensive and covers BDA events up to December 2013. Since then, there have been many relevant Big Data events.  Nevertheless, Press' review did cover both Big Data and Data Science events. To this extent, the term Data Science could be considered as a complementary meaning of BDA.

In comparison with Press' review, Frank Ohlhorst [5] established the origin of Big Data to 1880 when the 10th US census was held. The real problem during the 19th century was a statistics issue, which was how to survey and document 50 million of North-American citizens. Although Big Data may contain computation of some statistics elements, these two terms have different interpretations today. Similarly, Winshuttle [6] believe the origin of Big Data was in the 19th century. They argue if data sets are so large and so complex and beyond traditional process and management capability, then these data sets can be considered as "Big Data". In comparison to Press', Winshuttle's review emphasizes Enterprise Resource Planning (ERP) and implementation on cloud infrastructure. Moreover, the review also makes a predication for data growth to 2020. The total time span of its review was more than 220 years. Winshuttle's Big Data history included many SAP events and its data products, such as HANA.

The longest span of historical review for Big Data belongs to Bernard Marr's description [7]. He traced the origin of Big Data back to 18,000 BCE. Marr argued that we should pay attention to historical foundations of Big Data, which are different approaches for human to capture, store, analyze and retrieve both data and information. Furthermore, Marr believed that the first person who casted the term "Big Data" was Erik Larson [9], who presented an article for Harper's Magazine and it was subsequently reprinted in The Washington Post in 1989 because there were two sentences that consisted of the words of Big Data: "The keepers of Big Data say they do it for the consumer's benefit. But data have a way of being used for purposes other than originally intended."

In contrast, Steve Lohr [10] disagrees with Marr's view. He argues that just adopting the term alone might not have the today's Big Data connotation because "The term Big Data is so generic that the hunt for its origin was not just an effort to find an early reference to those two words being used together". Instead, the goal was the early use of the term that suggests its present interpretation — that is, not just a lot of data, but different types of data handled in new ways". This is an important point. Based on this reasoning, we consider that Cox and Ellsworth [8] as the origin of Big Data because they assigned a relatively accurate meaning to the existing view of Big Data, which they stated "…data sets are generally quite large, taxing the capacities of main memory, local disk and even remote disk. We call this the problem of Big Data. When data sets do not fit in main memory (in core), or when they do not fit even on local disk…". Although today's term may have extended meaning than Cox and Ellsworth's term, this definition reasonably accurately reflects today's connotation.

Another historical review was contributed by Visualizing .org [11]. It focused on the timeline of how to implement BDA. Its historical description is mainly determined by events related to Big Data push by many Internet and IT companies, such as Google, Youtube, Yahoo, Facebook, Twitter and Apple. Especially, it emphasized the significant impact of Hadoop in the history of BDA. It primarily highlighted the significant role of Hadoop in the BDA. Based on these studies, we show the history of Big Data, Hadoop and its ecosystem in Figure 1.

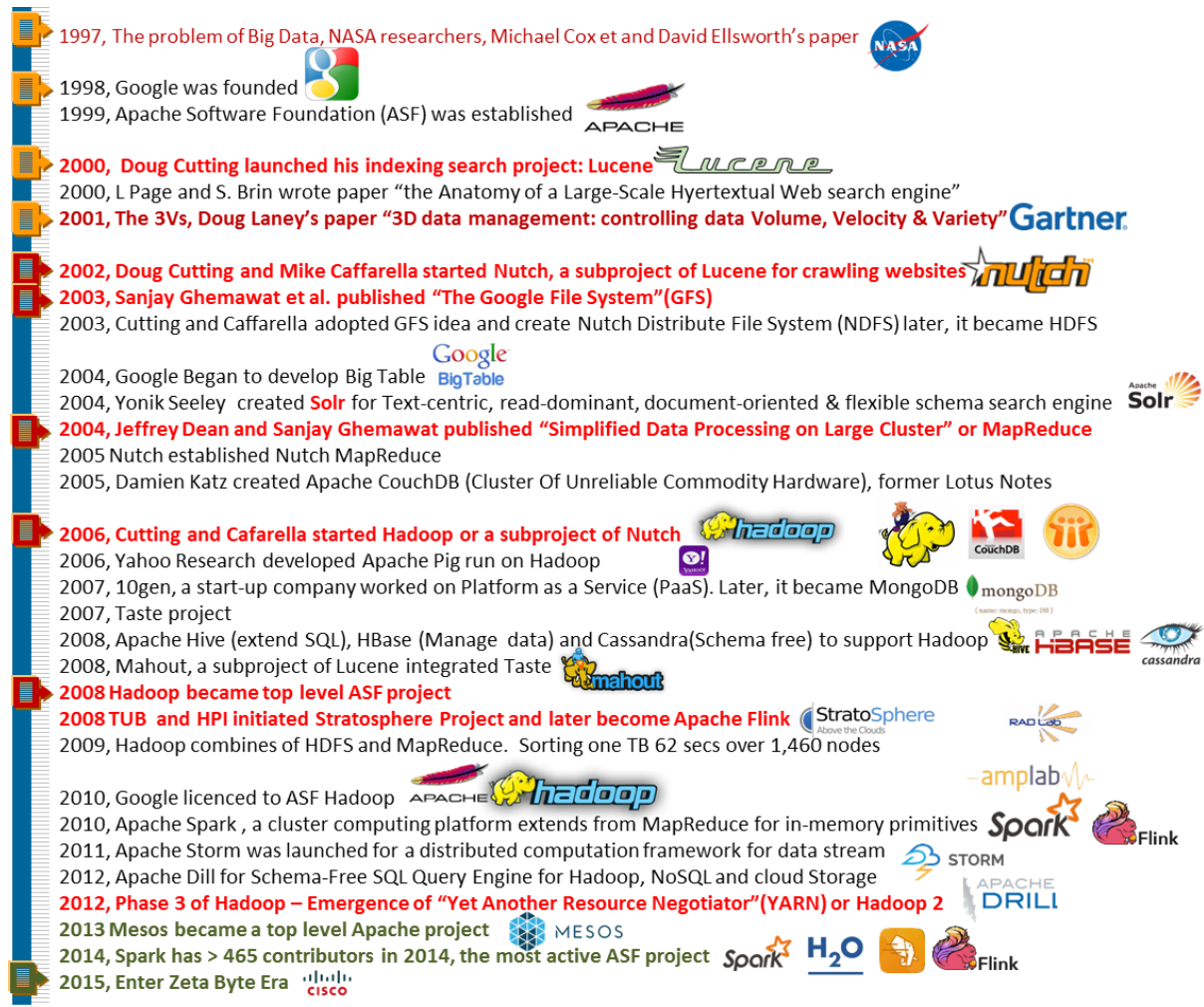

Figure 1 A Short History of Big Data

Undoubtedly, there will be many different views based on different interpretations of BDA. This will inevitably lead to many debates of Big Data implication or pros and cons.

### 1.2.2   Debates of Big Data Implication

*Pros*

There have been many debates regarding Big Data' implication during the past few years. Many advocates declare Big Data as a new rock star [20] and Big Data will be the next frontier [21], [22] for innovation, competition and productivity because data is embedded in the modern human being's life. Data that are generated by both machines and human in every second is a by-product of all other activities. It will become even the new epistemologies [23] in science. To certain degree, Mayer and Cukier [24] argued Big Data would revolutionize our way of thinking, working and living. They believe that a massive quantitive data accumulation will lead to qualitative advances at the core of BDA - machine learning, parallelism, metadata and predictions. "Big Data will be a source of new economic value and innovation". Their conclusion is that data can speak for itself and we should let the data speak.

To certain extent, Montjoye et al's [25] echoed the above conclusion. They demonstrated that it is highly probable (over 90% reliability) to re-identify a person with as little as only four spatiotemporal data points (credit card transactions in a shopping mall) by leveraging Big Data Analytics. Their conclusion is that "large scale data sets of human behavior have the potential to fundamentally transform the way we fight diseases, design cities and perform research."

*Cons*

In contrast, some argue that Big Data is inconclusive, overstated, exaggerated and misinformed by the media and data cannot speak for itself [12]. It does not matter how Big Dataset is. It could be just another delusion because "it is like having billions of monkeys typing, one of them will write Shakespeare." [13]. In Dobelli's term [14], we should "never judge a decision by its outcome – outcome bias". In other words, if one of the monkeys can type Shakespeare, we cannot conclude or inference that a monkey has sufficient intelligence to be Shakespeare.

Gary Drenik [15] believed that the sentiment of the overeager adoption of Big Data is more like "Extraordinary Popular Delusion and the Madness of Crowds", the description made by Charles Mackay [16] on his famous book's title. Psychologically, it is a kind of a crowd emotion that seems to have a perpetual feedback loop. Drenik quoted this "madness" with Mackay's warning: "We find that whole communities suddenly fix their minds upon one subject, and go mad in its pursuit; that millions of people become simultaneously impressed with one delusion, and run it till their attention is caught by some new folly more captivating than the first.". The issue that Drenik has noticed "the hype overtaken reality and there was little time to think about" regarding Big Data.    The former Obama's campaign CTO, Harper Reed, has the real story in terms of adoption of BDA. His remarks of Big Data were "literally hard" and "expensive" [34].

Danah Boyd et al [17] are quite sceptical regarding Big Data in term of its volume. They argued bigger data are not always better data from social science perspective. In responding to "The End of Theory" [18] proposition, Boyd asserted that theory or methodology is still highly relevant for today's statistical inference and "The size of data should fit the research question being asked; in some cases, small is best". They suggested that we should not pay a lot of attention to the volume of data. Philosophically, the critic is similar as the debate between John Stuart Mill (five Mill's classical or empirical methods) and his critics [35] in 19[th] century, which Mill's critics argued that it is impossible to bear on the intelligent question by just ingesting as much as data alone without some theory or hypothesis. This means that we cannot make Big Data do the work of theory.

Another Big Data critique comes from David Lazer et al. [19]. They demonstrated that Google Flu Trends (GFT) prediction is the parable and identified two issues (Big Data hubris and algorithm dynamics) that contributed to GFT's mistakes. The issue of "Big Data hubris" is that some observers believe that BDA can replace traditional data mining completely. The issue of "algorithm dynamics" is "the changes made by (Google's) engineers to improve the commercial service and by consumers in using that service". In another words, the changing algorithms for searching will directly impact on the users' behavior. This will lead to the collected data is driven by deliberated algorithms. Lazer concluded there are many traps in BDA, especially for social media research. Their conclusion was "we are far from a place where they (BDA) can supplant more traditional methods or theories."

All these multiple views were due to different interpretations of Big Data and different implementations of BDA. This suggests that in order to resolve these issues, we should first clarify the definition of the term BDA and then discover the clash point based on the same term.

# 1.3    Historical Interpretation of Big Data

### 1.3.1    Methodology for Defining Big Data

Intuitively, neither yesterday's data volume (absolute size) nor today's one can be defined as "Big". Moreover, today's "Big" may become tomorrow's "small". In order to clarify the term Big Data precisely and settle down the debate we can investigate and understand the functions of a definition based on the combination of Robert Baird's [26] and Irving Copi's [27] approaches (see Figure 2).

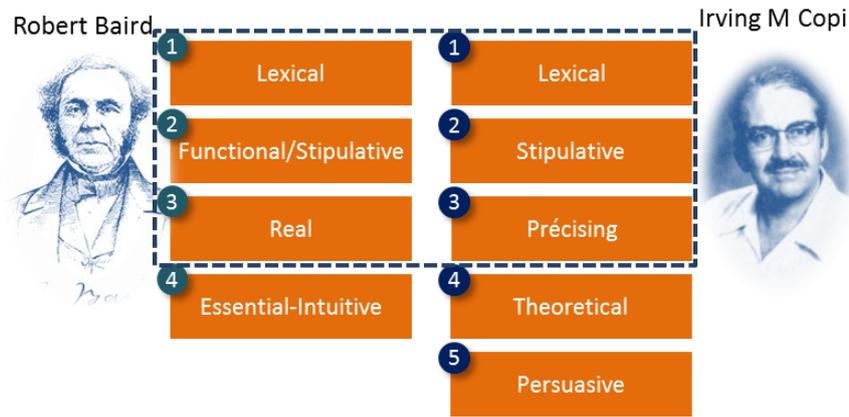

Figure2: Methodology of Definition

Based on Baird or Irving's approach of definition, we will first investigate the historical definition from an evolutionary perspective (lexical meaning). Then we extend the term from 3Vs to 9Vs or $3^2$ Vs based on its motivation (stipulative meaning), which is to add more attributes for the term. Finally, we will eliminate ambiguity and vagueness of the term and make the concept of Big Data more precise and meaningful.

### 1.3.2.  Different Attributes of Definitions

*Gartner- 3Vs Definition*
Since 1997, many attributes have been added to Big Data. Among these attributes, three of them are the most popular, which they have been widely cited and adopted: The first one is so called Gartner's interpretation or 3Vs. The root of this term can be traced back to February 2001. It was casted by Douglas Laney [28] in his white paper published by Meta group, which Gartner subsequently acquired in 2004. Douglas noticed that due to surging of e-commerce activities, data has grown along three dimensions, namely:

1.  Volume, which means Incoming data stream and Cumulative volume of data.
2.  Velocity, which represents the pace data used to support interaction and generated by interactions
3.  Variety, which signifies the variety of incompatible and inconsistent data formats and data structures.

According to the history of Big Data timeline [28], Douglas Laney's 3Vs definition has been widely regarded as the "common" attributes of Big Data but he stopped short of assigning these attributes to the term "Big Data".

*IBM- 4Vs Definition*
IBM added another attribute or "V" for "Veracity" on the top of Douglas Laney's 3Vs notation, which is so called as Four Vs of Big Data. It defines each "V" as following [29] [30]:

1.  Volume stands for scale of data
2.  Velocity denotes to analysing streaming data
3.  Variety indicates different forms of data
4.  Veracity implies uncertainty of data

Paul C. Zikopoulos et al. [31] explained the reason behind the additional "V" or veracity dimension, which is "*in response to the quality and source issues our clients began facing with their Big Data initiatives.*" They are also aware of some analysts including other V-based descriptors for Big Data, such as variability and visibility.

*Microsoft - 6Vs Definition*
For the sake of maximising the business value, Microsoft extended Douglas Laney's 3Vs attributes to 6 Vs [32], which it added Variability, Veracity and Visibility:

1.  Volume stands for scale of data
2.  Velocity denotes to analysing streaming data
3.  Variety indicates different forms of data
4.  Veracity focuses on trustworthiness of data sources.
5.  Variability refers to the complexity of data set. In comparison with "Variety" (or different data format), it means the number of variables in data sets.
6.  Visibility emphasise that you need have a full picture of data in order to make informative decision.

*More Vs for Big Data*

There has been also a 5 Vs Big Data definition presented by Yuri Demchenko [33] in 2013. He added the value dimension along with the IBM 4Vs' definition (see Figure 3). Since Douglas Laney published 3Vs in 2001, there have been many additional "Vs". We can find the number of Vs as many as eleven [41].

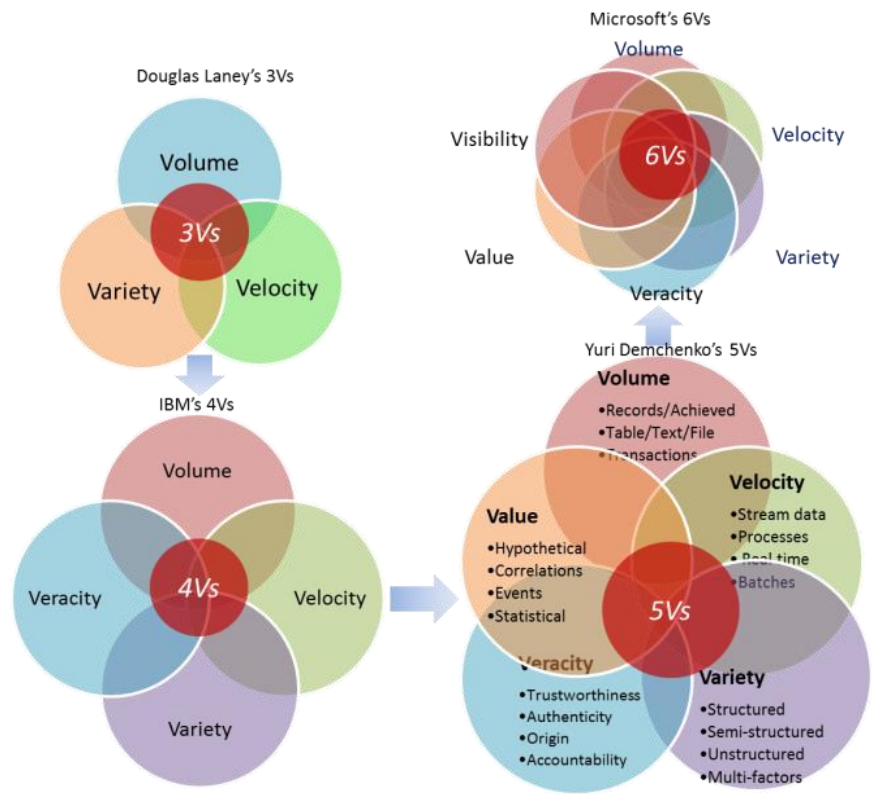

Figure 3: From 3Vs, 4Vs, 5Vs and 6Vs Big Data Definition

All these definitions, such as 3Vs, 4Vs, 5Vs or even 11 Vs are primarily trying to articulate the aspect of data. Most of them are the data-orientated definitions but fail to articulate Big Data clearly in a relationship to the essence of BDA. In order to understand the essential meaning, we have to clarify what data is.

Data is everything within the universe. This means that data is within the existing limitation of technological capacity. If the technology capacity is allowed, there is no boundary or limitation for data. The question is why we should capture it in the first place. Clearly, the primary reason of capturing data is not because we have the capacity to capture high volume, high velocity and high variety data rather than expect to find a better solution for our research or business problem, which is to search for actionable intelligence. Pure data driven analysis may add little value for a decision maker. Sometime, it may only add the burden for the costs or resources of BDA. Perhaps, this is why Harper believes Big Data is really hard [34].

### 1.3.3 Summary of 7 types Definitions of Big Data

Table 2 shows seven types of definitions, summarized by Timo Elliott [41], based on more than 33 Big Data definitions [42].

| No | Type | Description |
|---|---|---|
| 1 | The Original Big Data (3Vs) | The original type of definition is referred to Douglas Laney's Volume, Velocity and Variety or 3Vs. It has been widely cited since 2001. Many have tried to extend the number of Vs, such as 4Vs, 5Vs, 6Vs … up to 11 Vs |
| 2 | Big Data as Technology | This type of definition is oriented by new technology development, such as MapReduce, Bulk Synchronous Parallel (BSP - Hama), Resilient Distributed Datasets (RDD, Spark), and Lambda architecture (Flink). |
| 3 | Big Data as Application | This kind of definition emphasizes different applications based on different types of Big Data. Barry Devlin [43] defined it as application of process-mediated data, human-sourced information and machine generated data. Shaun Connolly [44] focused on analyzing |
| | | transactions, interactions and observation of data. It looks for hindsight of data |
| 4 | Big Data as Signals | This is another type of application oriented definition but it focuses on timing rather than type of data. It looks for a foresight of data or new 'signal' pattern in dataset |
| 5 | Big Data as Opportunity | Matt Aslett [45]: "Big Data as analyzing data that was previously ignored because of technology limitations". It highlights many potential opportunities by revisiting the collected or archived datasets when new technologies are variable. |
| 6 | Big Data as Metaphor | It defines Big Data as human thinking process [46]. It elevates BDA to the new level, which BDA is not just a type of analytics rather than the extension of human brain. |
| 7 | Big Data as New Term for Old Stuff | This definition simply means the new bottle (relabel the new term "Big Data") for old wine (Business intelligence or data mining or other traditional data analytic activities). It is one of the most cynical ways to define Big Data |

Table 2: Seven Popular Big Data Definitions

Each of the above definitions intends to describe a particular issue from one aspect of Big Data only and is very restrictive. However, a comprehensive definition can become complex and very long. A solution for this issueis to use "rational reconstruction" offered by Karl Popper, which intends to make the reasons behind practice, decision and process explicit and easier to understand.

### 1.3.4 Motivations behind the Definitions

The purpose of doing Big Data or BDA is to gain hindsight (metadata patterns emerging from historical data), insight (deep understanding of issues or problems) and foresight (accurate prediction in near future) in a cost effective manner. However, these important and necessary attributes are often neglected by many definitions that only focus on either single issue or data aspects. In order to reflect all aspects of Big Data, we consider all attributes from different aspects.

# 1.4 Defining Big Data from 3Vs to $3^2$ Vs

The real objective of BDA is actually to seek for Business Intelligence (BI). It enables decision makers to make right decision based on predictions through the analysis of available data. Therefore, we need to clarify new attributes of Big Data and establish their relationship meaning cross three aspects (or domain knowledge), namely:

• Data Domain (Searching for patterns)
• Business intelligent Domain (Making predictions)
• Statistical Domain (Making assumptions)

### 1.4.1 Data Domain

Laney's 3Vs have captured importance of Big Data characteristics reflecting the pace and exploration phenomena of data growth during the last few years. In this, the key attribute in data aspect is Volume. If we look the history of data analytics, the variation of velocity and variety is relatively small in comparison with volume. The dominated "V" that is often exceeds our current capacity for data processing is "Volume". Although volume cannot determine all attributes of data, it is one of the crucial factors in BDA.

### 1.4.2 Business [1] Intelligent (BI) Domain

When we discuss BI of BDA, we mean Value, Visibility and Verdict within the business intelligent domain. These 3Vs are the motivations or drivers for us to implement BDA process at the first place. If we cannot achieve BI, the pure exercise of data analytics will be meaningless. From a decision maker's perspective, these 3Vs are how to leverage Data's 3Vs for BI's 3Vs.

• Visibility: it does not only focus on the insight but also means metadata or sometime the wisdom of data crowds or hierarchical level of abstraction data patterns. From BI perspective, it provides hindsight, insight and foresight of a problem and an adequate solution associated with it.
• Value: the purpose of V for value is to answer the question of "Does the data contain any valuable information for my business needs?" In comparison with 5Vs definition, it is not just the value of data but also the value of BI for problem solving. It is the value and utility for the long term or strategic pay off.
• Verdict: It is a potential or possible choice or decision should be made by a decision maker or decision committee based on a scope of problem, available resources and certain computational capacity. This is the most challenging V to be quantified at the beginning of BDA. If there are many hypothesise of "What-if"s, the cost of collecting, retrieving data, ETL, especially to extract archived data would be costly (see Figure 4).

---

[1] Here, the term of business includes research activities.

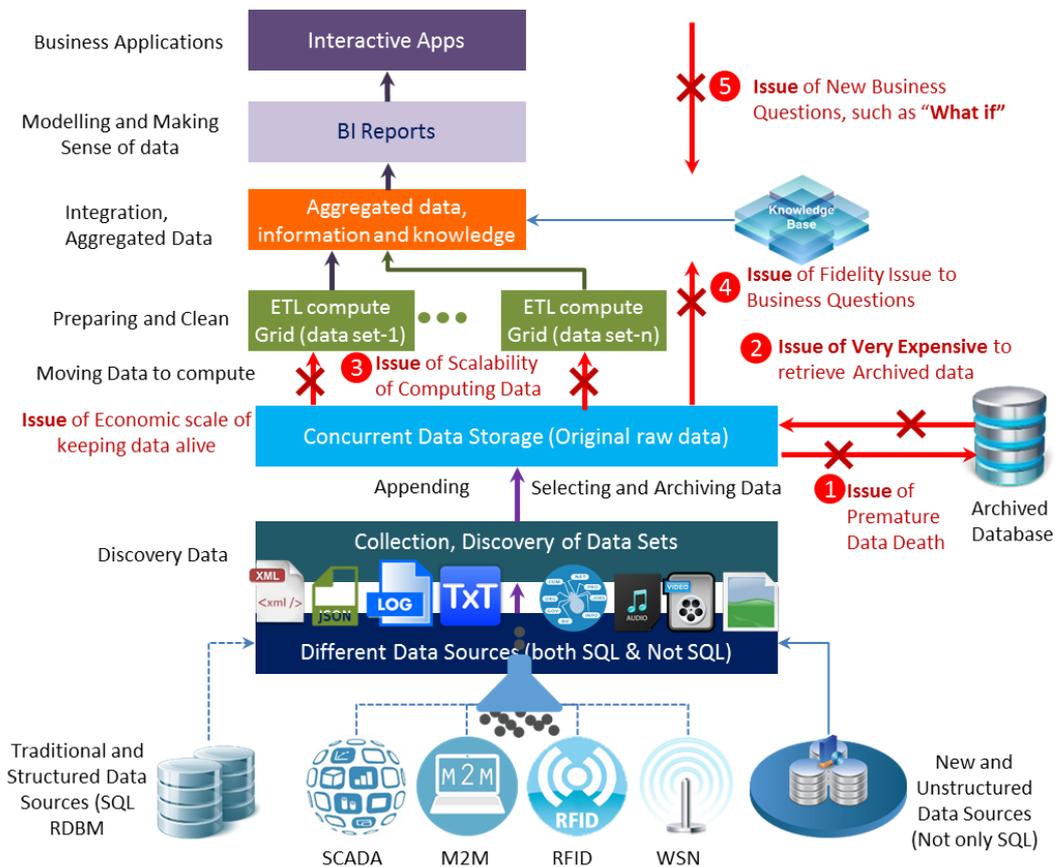

Figure 4: Key Motivations of Big Data Analytics

These business motivations led to the new BDA platforms or MapReduce processing frameworks such as Hadoop. It intends to answer the five basic questions in Big Data as shown in Figure 4. These questions reflect the bottom line of Business Intelligence (BI):

1. How to store massive data (such as in PB or EB scale currently) or information in the available resources
2. How to access these massive data or information quickly
3. How to work with datasets in variety formats: structured, semi-structured and unstructured
4. How to process these datasets in full scalable, fault tolerant and flexible manner
5. How to extract business intelligence interactively and relational way in a cost effective manner

In this domain, the key notation of V is "Visibility", which is to obtain the prediction or real time insight from BDA exercises. The relationship of these 3Vs in BI is that without visibility, other 2Vs will be impossible.

### 1.4.3 Statistics Domain

Similarly, we should have another set of 3Vs attributes in the statistic domain, which are Veracity, Validity and Variability. These 3Vs should establish the statistic models based on right hypothesis (What–if), which is the trustworthiness of data sets' and the reliability of data sources. If the hypothesis is inadequate or the data source is contaminated or the statistics model is incorrect, the BDA might lead to a wrong conclusion. There have been many lessons regarding contaminated data samples. A famous example was the opinion poll for the 1936 US presidential election that was carried by the Literary Digest magazine before the election [36]. Because the sample data (2.4 million survey responses) was accidentally contaminated, the result of its predication (or president winner in 1936) became a disaster for the polling company. Attributes in this domain are:

- Veracity: Philosophically speaking, the true information (or fact) is the resolution of data uncertainty. V of Veracity is searching for trustworthiness and certainty of data sets.
- Validity: It is to verify the quality of data being logically sound. The V of validity emphasizes how to correctly acquire data and avoid biases. Another essential meaning of validity is the inference process based on a statistical model.
- Variability: It is the implication of data complexity and variation. For example, Bruce Ratner [37] believed that if there are more than 50 variables or different features in one dataset, it could be considered as "Big Data". Statistically, it is how to use the logical inference process to reduce data complexity and reach desirable outcomes or predictions for business needs.

The key attribute of this aspect is "Veracity", which emphasizes how to build a statistical model close to the reality. The process to approach "Veracity" can be considered as an exercise of a curve fitting. If we have few constraints, the regression errors of the curve will be too large. If we adopt too many constraints, it will cause an over-fitting problem.

### 1.4.4 $3^2$ Vs Definition and Big Data Venn Diagram

Once all $3^2$ Vs attributes have been defined from three different aspects, we can establish a combined Venn diagram and their relationships. This has become our definition of Big Data (see Figure 5), which is comprehensive enough to capture all aspects of Big Data.

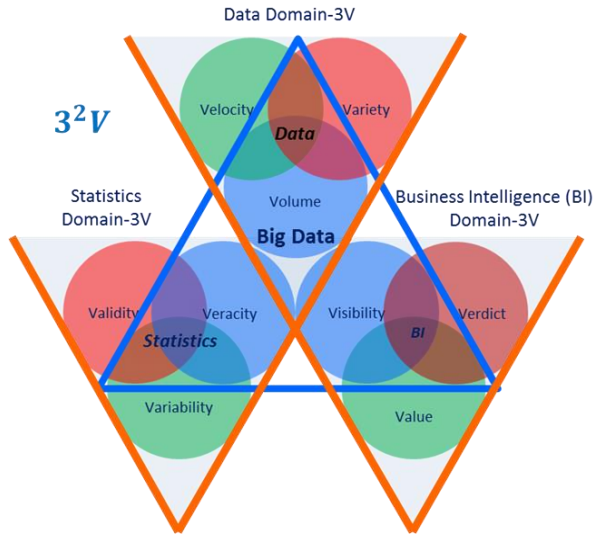

Figure 5: $3^2$ Vs Venn Diagrams in Hierarchical Model

As shown Figure 5, each Venn diagrams is supported by one "V" shape of triangle to illustrate 3 Vs attributes in one aspect. Moreover, three key attributes from each Venn diagram can also form a single hierarchical triangle diagram. It represents the essential meaning of Big Data.

If the original 3Vs data attributes represented a syntactic or logical meaning of Big Data, then $3^2$ Vs (or 9Vs) represent the semantic meaning (relationship of data, BI and statistics). For many complex problems or applications, the $3^2$ Vs could be interpreted as a hierarchical model, which three key attributes forms a higher level 3Vs to be learnt by a machine. At the heart of BDA, there is "machine learning" because without the machine (computer), the mission of learning from Big Data would be impossible.

# 1.5 Big Data Analytics and Machine Learning

### 1.5.1 Big Data Analytics

If $3^2$Vs represent semantic meaning of Big Data, then Big Data Analytics (BDA) represents pragmatic meaning of Big Data. We can view from computational view point, Big Data Venn diagram with a BDA's Venn diagram in Figure 6.

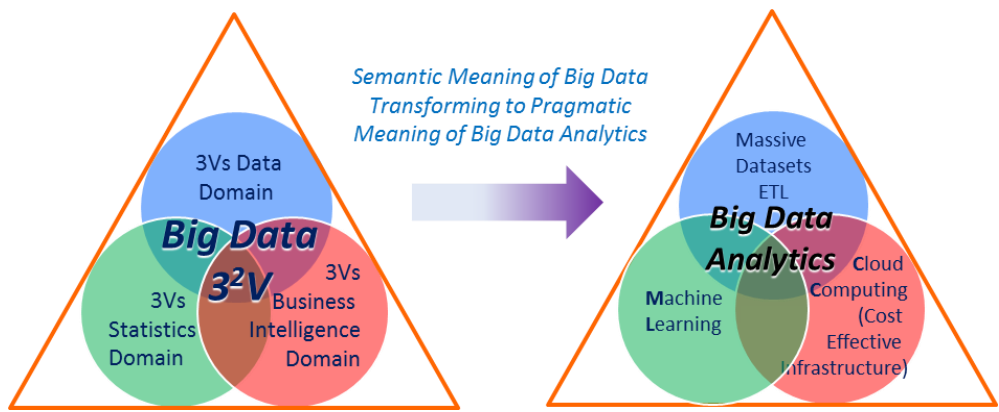

Figure 6: Correlation of $3^2$ Vs to Machine Learning Venn Diagrams

According to Arthur Samuel, the original definition of Machine Learning (ML) was *"The field of study that gives computers (or machine) that ability to learn without being explicitly programmed"* [38]. Historically, there have been many terms that intend to describe the equivalent meaning of ML, such as "Learning from data", "Pattern Recognition", "Data science", "Data Mining", "Text Mining" or even "Business Intelligence" and etc. If we list all terms based on their different orientations, we can probably find there are more than 32 different descriptions that contain certain meaning of ML from four aspects (see Table 3):

- Data
- Information
- Knowledge and
- Intelligence

| Data | Information | Knowledge | Intelligence |
|------|-------------|-----------|--------------|
| Data Mining | Information Analytics | Real time Analytics | Business analysis |
| Data Science | Information visualization | Predictive analytics | Business Intelligence |
| Data Warehouse | Information System Management | Machine Learning | Artificial Intelligence |
| Learning from Data | Text Analytics | Knowledge Base System | Decision Support System |
| Data Smart | Text Mining | Pattern Recognition | Actionable Intelligence |
| Data Analytics | Web Analytics | Statistical Application | Business Forecasting |
| Making Sense of Data | Web Semantic Analysis | Knowledge Discovery | Business Strategy |
| Data Ingestion | Web Searching | Expert Systems | Business Transformation |

Table 3: Popular Interpretation of ML

### 1.5.2 Machine Learning

The essence of ML is an automatic process of pattern recognition by a learning machine. The main objective of machine learning is build systems that can perform at or exceed human level competence in handling many complex tasks or problems. Machine learning is a part of Artificial Intelligence (AI). During the early AI research era, the AI's goal was to build robots and to simulate human activities. Later, the application of AI has been generalized to solve general problems by a machine. The popular solution was to feed a computer with algorithms (or a sequence of instructions) so it can transform the input data to output answers. This is often called as a rule based system or Good Old Fashion of AI (GOFAI), such as expert systems.

However, for many problems, we cannot easily find suitable algorithms, for example, the recognition of human handwriting. We do not know how to transform the input of hand writing letter to the output of the standard recognised letter. An alternative is learning from data. The principle of learning from data is similar as both trial-error and "The Wisdom of Crowds" [40]. This means that having one trial, it has a large error but if we can aggregate many trials, the error will be reduced down to an acceptable level or convergence. Figure 7 illustrates a typical example of machine learning process or learning from data.

Since the dotcom boom started in late 1990s, the volume of data has become increasingly larger. A logical question is how to deal with these large volumes of data and how to find useful or meaningful patterns from larger volume of data. This leads to "knowledge discovery in database" (or KDD), which is also called as data mining. In other words, we want to dig in the database and discover the meaning or knowledge for decision making. Larose et al. [47] defined the term as "the process of discovering useful patterns and trends in large datasets". In order to discover meaningful patterns from massive data set, statistics is the vital tool to add the value for data sampling, modelling, analysis, interpretation and presentation just as Jiawei Han et al. [48] indicated, *"Data mining has an inherent connection with statistics"*. This leads to converging of data mining and fuzzy expert system under the big umbrella of machine learning. From machine learning evolution perspective, the statistics theory or probability modelling has shifted AI discipline from rule-based expert systems or schema-on-write learning to a schema-on-read or data-driven methodology, which is to resolve the uncertainty issue with parameters' probability of a model. From this perspective, the statistics has been embedded into machine learning. As Witten et al [49] indicated, *"In truth, you should not look for a dividing line between machine learning and statistics because there is a continuum — and a multidimensional one at that—of data analysis techniques."*

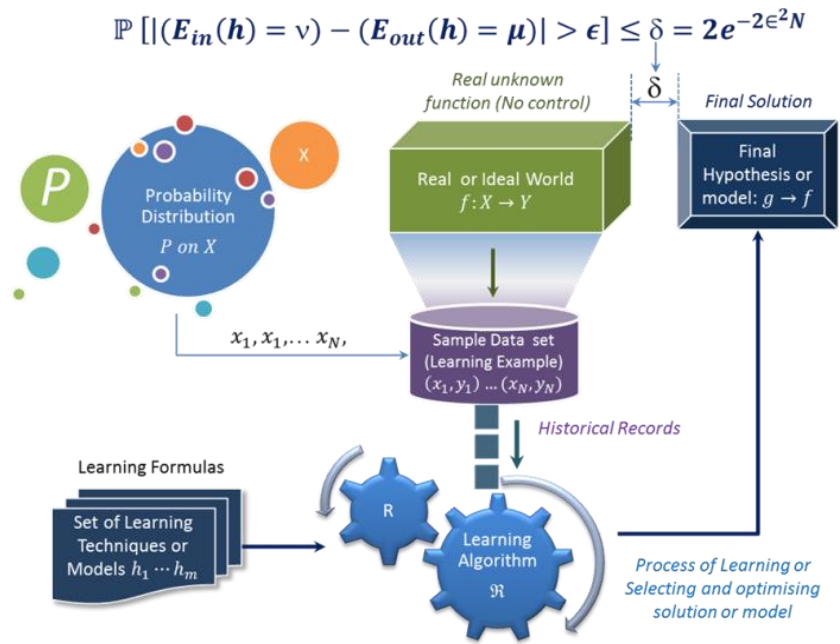

$$\mathbb{P}\left[\left|(E_{in}(h) = v) - (E_{out}(h) = \mu)\right| > \epsilon\right] \leq \delta = 2e^{-2\epsilon^2 N}$$

Figure 7: Machine Learning Process [39]

Since 1950s, there have been many functional definitions of ML. Different authors would emphasize different aspects of machine learning, such as process, application and utility. For example, Arthur Samuel's definition emphasized on "automatically learning" of ML. Tom M. Mitchell described every component of ML process [50]. Kevin P. Murphy [51] and Christopher M. Bishop [52], on the other hand, stressed the function of pattern recognition. Noam Nisan and Shimon Schocken [53] argued that ML could turn abstract thoughts into physical operation. In the summary of over 30 definitions, we can find some of essential and common ingredients of these ML definitions:

• Train the machine to learn automatically and improve results as it gets more data
• Discovery or recognize patterns and intelligence with input data
• Predicate on unknown inputs
• Machine will acquire knowledge directly from data and solve problems

According to these elements, we can find that fundamentally, ML is "*an outgrowth of the intersection of computer science and statistics, aims to automatically learn to recognize complex patterns and make intelligent decisions based on existing datasets*" [54]. Another way to say that is "*Machine learning is turning data into information*" [55]. The ultimate goal of ML is to build systems that are of level of human competence (see Figure 8) in performing complex tasks.

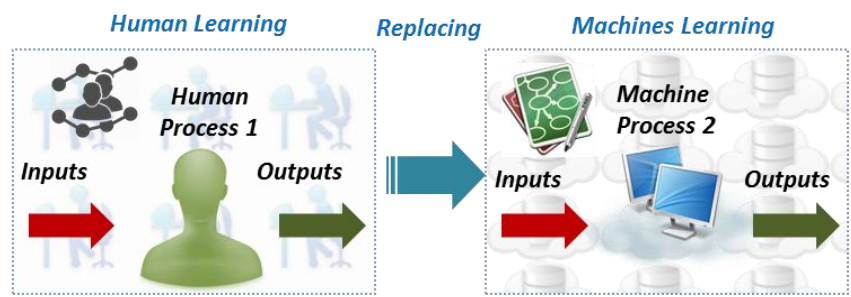

Figure 8: Replacing Human in the Learning Process

ML underpins the BDA implementation. If without ML to mine ever-growing massive data, BDA would be impossible. In conclusion, ML is the centrepiece of any BDA. All other components within a framework of Big Data aim to support ML process. In terms of computational support to BDA, there are four major architectural models that are able to process large amounts of data in a reasonable time according to S. Wadkar et al [56]:

• Massively parallel processing (MPP) database system: For example, EMC's Greenplum and IBM's Netezza
• In-memory database systems such as Oracle Exalytics, SAP's HANA and Spark

- MapReduce processing model and platforms such as Hadoop and Google File System (GFS)
- Bulk Synchronous Parallel (BSP) systems such as Apache HAMA and Giraph

To perform BDA in the most cost effective way, a fifth model—Cloud Computing (CC) — has become a preferred solution especially for small and media businesses (SMEs).

# 1.6 Big Data Analytics and Cloud Computing

Cloud Computing (CC) plays a critical role in BDA process as it offers subscription-oriented access computing infrastructure, data, and application services [74]. The original objective of BDA was to leverage commodity hardware to build computing clusters and scale out the computing capacity for web crawling and indexing system workloads. Due to the massive volume of dataset, searching for lower cost and fault tolerance computational capacity is an important factor for implementing BDA. On the other hand, the implementation of cloud computing were underpinned with 3 service models, 4 deployment models and – 5 Characteristics [76], which is so called 3S-4D-5C definition.

- o Service orientation or 3 S Service models (SaaS, PaaS and IaaS)
- o Customized delivery or 4D Deployment models (Private, Public, Community and Hybrid Cloud)
- o Shared Infrastructure or 5C Characteristics (On-Demand, Broad network Access, Resource Pool, Rapid elasticity and measured service)

This means that the nature of cloud characteristics makes it as the most accessible infrastructure for many small to medium companies to be able to implement BDA.

Cloud does not only enable us to easily scale out but also scale down to fit all sizes of dataset. When BDA is discussed, it is quite often that the only focus is how to scale out. However, it is not a necessary the case. Although the overall data volume may trend to increase, the daily volume for each individual case could be moderate and fluctuating or Big Data processing requirements needed for BI can vary from time to time. If we can leverage the elastic nature of cloud, we can save substantial amount of cost due to amortization benefits provided by the Cloud systems. The elastic nature of cloud can reduce the overall cost of computation for different types of Big Data workloads, such as batch, micro-batch, interactive, real time, and near real time.

Taking Yahoo sorting one TB data as an example, it took 3.5 minutes over 910 nodes to complete the task in 2008 but it only took 62 seconds over 1,460 nodes in 2009. To scale-out computational capacity did make huge difference regardless an improvement of each node due technological advances. This implies that cloud infrastructure provides computational flexibility if Big Data workload or business requirements need. For example, Amazon Web Service (AWS) offers spots instances at a fraction of the regular rate. If the workload only requires batch mode, we can leverage AWS's spots instance to increase computational capacity and complete the job in a much shorter time.

A popular and open platform that is widely deployed on a cloud infrastructure is Hadoop, whose implementation is inspired by Google MapReduce and Google File System (GFS).

# 1.7 Hadoop, HDFS, MapReduce, Spark and Flink

Figure 9 highlights one of the most popular platforms of BDA - Hadoop. It was the first choice for many analysts and decision makers for implementing BDA. One of the two Hadoop's founders - Michael Cafarella remarked, "Nutch (the predecessor of Hadoop) is The National Public Radio (NPR) of search engines" [63]. There are several reasons behind this development:

1. It is an open source platform and also programmed in java.
2. It is linearly scalable, reliable and accepts hardware failure.
3. It is a fault tolerant system
4. It is a practical platform to store and process greater than 10s of TB data
5. It leverages commodity type of hardware
6. It is "schema on read" or has "data agility" character
7. It is best fit for diversified data sources

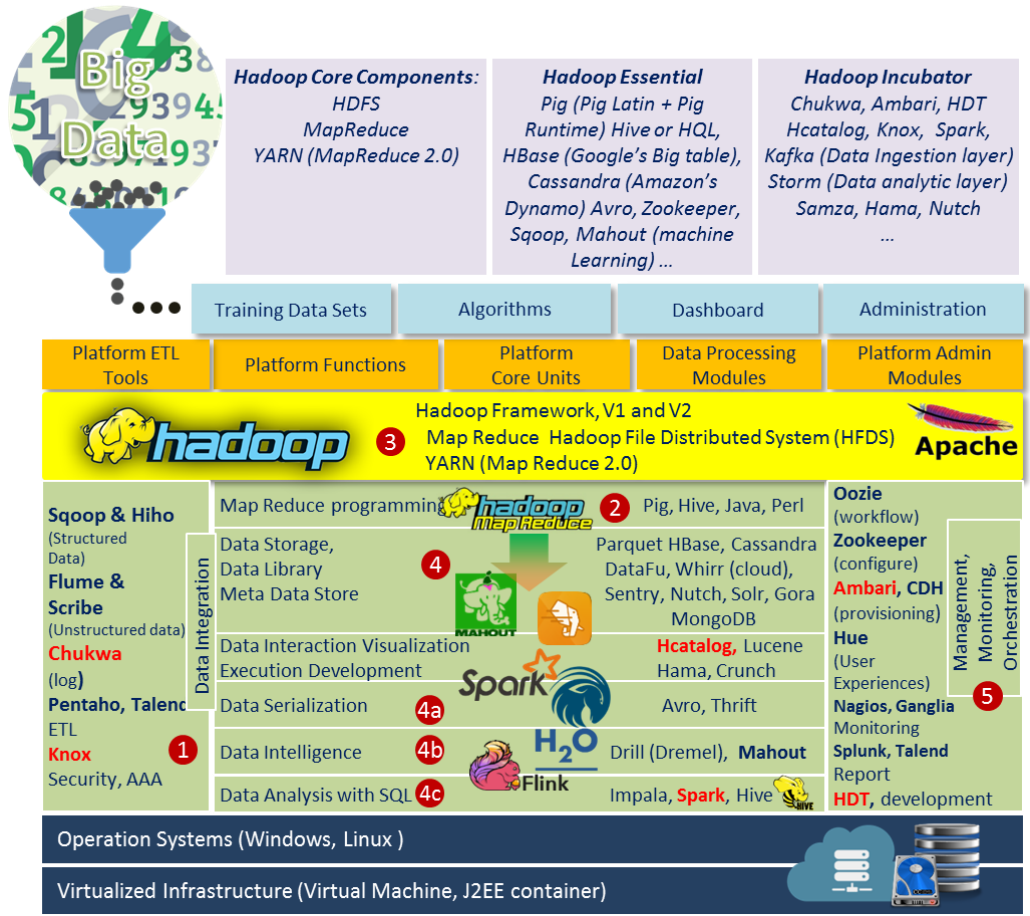

Figure 9: Overview of Hadoop Framework or Technology Stack and Ecosystem

The basic idea to create Hadoop is driven by both ever-growing data and cost of computational hardware. The objective of Hadoop is to leverage the commodity hardware for large scale of processing workload, which it used to be only possible to be accomplished by some expensive mainframe computers. From an infrastructure perspective, Hadoop enables the computational capacity to be scale-out rather than scale-up. Notice that it is quite often we use both terms interchangeably [57] but based on a standard definition, "scale-up" has a quality improvement sense while "scale-out" implies adding or repeating the same unit horizontally.

The advantage to adopt Hadoop [57] platform is that "*Hadoop is a free and open source distributed storage and computational platform. It was created to allow storing and processing large amounts of data using clusters of commodity hardware*". This statement also describes the basic principle of Hadoop architecture that consists of three essential components (see Figure 10): Hadoop Distributed File System (HDFS) for file storage function, Map for distribute function and Reduce for parallel processing function.

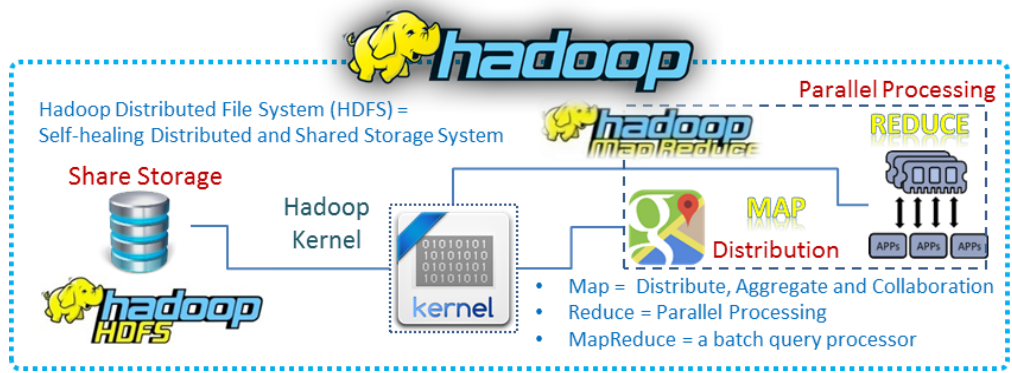

Figure10: Hadoop Kernel

However, its main disadvantage is that it processes all workloads in batch mode because *"Hadoop is a generic processing framework designed to execute queries and other batch read operations on massive datasets that can scale from tens of terabytes to petabytes in size"*[58]. This means that the early version of Hadoop cannot handle streaming and interactive workloads. Table 5 summarizes main characteristics of Hadoop.

| Attributes | Characteristics of Hadoop |
|---|---|
| Initiators | Doug Cutting and Michael J Cafarella |
| Predecessor | Nutch |
| Subsequent Version | YARN or Hadoop 2.0 |
| Hadoop Written Language | Java |
| Philosophy of computation | Divide and Conquer for large datasets |
| Principle of Computational Processing | Bring computer to data rather than bring data to computer |
| System | A distributed programming framework |
| Main Characteristics | Accessible, Robust, Scalable, Simple and Fault tolerance |
| Storage -Hadoop Distributed File System (HDFS) | Self-healing Distributed and shared storage element |
| Initial Computational Program - MapReduce | Distributed, aggregated and collaborated parallel processing |
| MapReduce Library written language | C++ code |
| Process Type | Batch |
| Hardware Type | Heterogeneous commodity hardware |
| Software licence | Open Source |
| Initial Applications | Information Retrieval (IR) and searching index and Web Crawler |
| Solution Type | Software solution not hardware solution |
| Scalability Solution | Scale-out not Scale-up |
| Typical Size of Data Set | From few GBs to few TBs |
| Capable Size of Data Set | From Tens of TBs to Few PBs |
| Simple Coherency Model | Write-once and Read many |
| Default Replication Factor | 3 |
| A typical size of data block for HDFS | 64MB |
| Permission Model | Relaxing POSIX [2] model |
| Main Application Modules | Mahout, Hive, Pig, HBase, Sqoop, Flume, Chukwa, Pentaho … |
| Typical Vendors | MapR, Cloudera, Hortonworks, IBM, Teradata, Intel, AWS, Pivotal Software and Microsoft |

Table 5: Common Aspects of Hadoop

The origin of Hadoop can be traced back to Nutch project under Apache Software Foundation (ASF) in 2002 (see Figure 11). The initial platform was built as an open source implementation of MapReduce [60] processing model and distributed file system [60] proposed by Google. In 2010, Google has granted a license to Apache for incorporating MapReduce model in Hadoop software freely and distributed it without any patent or IP rights infringement concerns.

---

[2] POSIX = the Portable Operating System Interface. Few POSIX rules (permissions model for supporting multiuser environment) have been relaxed in order to gain higher throughput of data uploads.

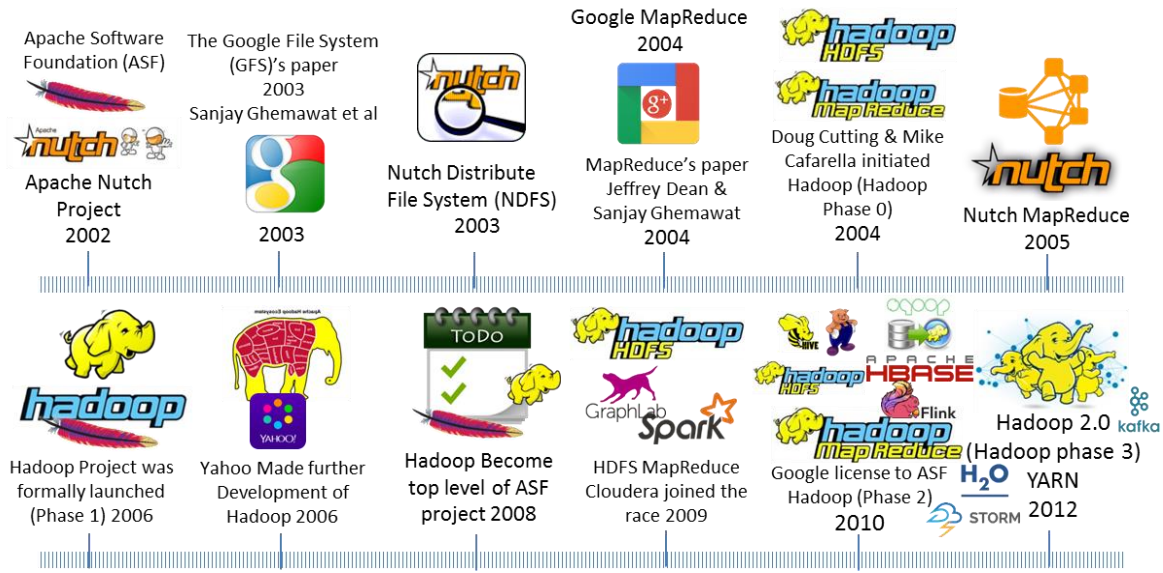

Figure 11: Briefing History of Hadoop

### 1.7.1    Google File System (GFS) and HDFS

The Hadoop project adopted "Google File System (GFS) architecture and developed Hadoop Distributed File System (HDFS). The original authors (Google's engineers) laid out four pillars for GFS:

- System Principles
- System architecture
- System assumptions and
- System Interfaces

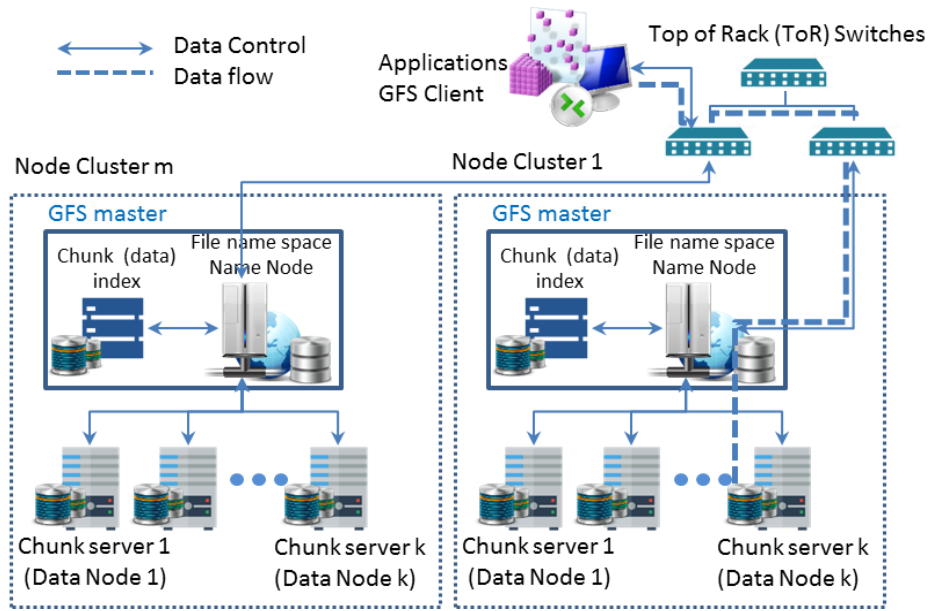

Figure 12: GFS or HDFS Architecture

The GFS principles departed from the traditional system design dogma that a failure was not allowed and a computation system should be designed as reliable as possible. In contrast, GFS anticipates the certain number of system failures with specified redundancy or replicating factor and automatic recovery. In comparison with the traditional file standard, GFS is capable of handling billions objects and I/O should be revisited. Moreover, most of files will be altered by appending rather than overwriting. Finally, the GFS flexibility is increased by balancing the benefits between GFS applications and file system API. The GFS architecture consists of three components (see Figure 12):

- Single master server (or name node)

- Multiple chunk servers (or data nodes for Hadoop)
- Multiple clients

The master server maintains 6 types of the GFS's metadata, which are: 1) Namespace, 2) Access control information, 3) Mapping from files to chunks (data), 4) Current locations of chunks or data, 5) System activities: (chunk lease management, garbage collection of orphaned chunks and chunk migration between chunk servers), 6) Master communication of each chunk server in heart beat messages.

GFS was designed with five basic assumptions [59] according to its particular application requirements:

1.  GFS will anticipate any commodity hardware outages caused by both software and hardware faults. This means that an individual node may be unreliable. This assumption is similar to one of its system design principles.
2.  GFS accepts a modest number of large files. The quantitive of "modest" is few million files. A typical file size is 100 MB/per file. The system also accepts smaller file but will not optimize them.
3.  Typical workload size for stream reading would be from hundred KBs to 1MB with small random reads for few KBs in batch mode
4.  GFS has its will defined sematic for multi-clients with minimal synchronization overhead
5.  A constant high file storage network bandwidth is more important than low latency

In contrast to other file systems, such as Andrew File System (AFS) or Serverless File System (xFS) or Swift, GFS does not adopt a standard API POSIX permission model rather than relax its rules to support the usual operations to create, delete, open, close and write.

According to these workload processing assumptions, GFS is actually a file storage system or framework that has two basic data structure: logs (metadata) and Sort String Table (SSTable). The main object of having GFS is to implement Google's data-intensive applications. Initially, it was designed to handle the issues of web crawler and file indexing system under the pressure of accelerating data growing.

The aim that Google published these influential papers [59] was to show how they scale out the file storage system for large distributed data-intensive applications. Doug Cutting and Mike Cafarella leveraged the Google's GFS idea to develop their file system – Nutch or Nutch Distribute File System (NDFS) for web crawling application, namely Apache Lucene. NDFS was the predecessor of HDFS (see Figures 13 and 15). Although HDFS is based on GFS concept and has many similar properties and assumptions as GFS, it is different with GFS in many ways, especially in term of scalability, data mutability, communication protocol, replication strategy, and security.

### 1.7.2 MapReduce

MapReduce is a programming model to process large dataset workload. In contrast to imperative programming (describing computation as a bunch of statements to change program state), MapReduce treats computation as the evaluation of mathematic function. In essence, functional programming can avoid state and just list in and out states.

The basic strategy of MapReduce is "Divide and Conquer". In order to perform different data intensive applications effectively with MapReduce on the GFS framework, Dean and Ghemawat [60] presented a five-step process (a programming model can be considered as a process, see Figure 13).

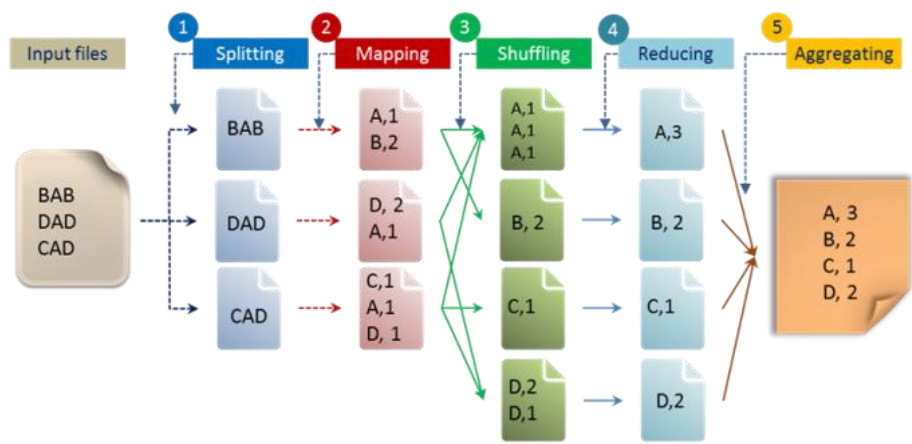

Figure 13: Five Steps MapReduce Programming Model

Step 1:  Splitting
Step 2:  Mapping (distribution)
Step 3:  Shuffling and sorting
Step 4:  Reducing (parallelizing)
Step 5:  Aggregating

Lin et al [62] simplified this process model into three steps, mapping, shuffling and reducing. As shown in Figure 13, the 1st step is the process of splitting input file into three files and the 2nd step is to generate a process of key/value pair by a user (or client) who specifies the function. In the above example, it is to count the number of different letters (A, B, C and D) with corresponding quantity within each split file. The 1st split file contains word "BAB". The letter "A" is counted 1 and letter "B" is counted 2. In the 3rd step, the shuffling function is to generate intermediate key/value pair, which is to sort the same letter (or key) and quantity (or value) from different split files into one file. The 4th step is to merge all intermediate values (3, 2, 1, and 2) associated with the same intermediate key (A, B, C, and D). The final step aggregates these key/value pairs into one output file. Here, "key" is equal to different types of letters to be counted and "value" is equal to the quantity of each letter.

From a programming perspective, MapReduce has other two meanings that "Mapping" is splitting for distribution and "Reducing" is shuffling + sorting in parallel. A major advantage is its capability of shared-nothing data processing, which means all mappers can process its data independently.

The characteristic of shared-nothing enable MapReduce to run a simple program cross thousands or even millions of unreliable and homogeneous machines in parallel and complete a task in very short time. Theoretically speaking, it allows any programmer to access almost unlimited commodity type of computing resources instantly (theoretically) or within an acceptable time frame (practically) e.g. cloud infrastructure. Several Cloud computing platforms have implemented their own MapReduce processing model such as CouchDB, Cloud MapReduce and Aneka [74].

According to Dean and Ghemawat [60], the original Google's MapReduce is potentially capable to handle five types of workloads:

1.  Large-scale machine learning problems,
2.  Clustering problems for the Google News and Google products,
3.  Extraction of data used to produce reports of popular queries (e.g. Google Zeitgeist),
4.  Extraction of properties of web pages for new experiments and products (e.g. extraction of geographical locations from a large corpus of web pages for localized search), and
5.  Large-scale graph computations

Eric Bieschke' echoed this point and indicated, "Hadoop is cost efficient, but more than that, it makes it possible to do super large-scale machine learning" [61]. To this extent, the history of Hadoop is an evolutionary progress to generalize data processing task from a particular workload (e.g. web crawler) to all types of ML workloads (see Figure 14). However, MapReduce is not very efficient to perform iterative and recursive process that is widely utilised for a simulation type of workload in ML. In order to understand the issue, it is necessary to see how the Hadoop project has been evolved.

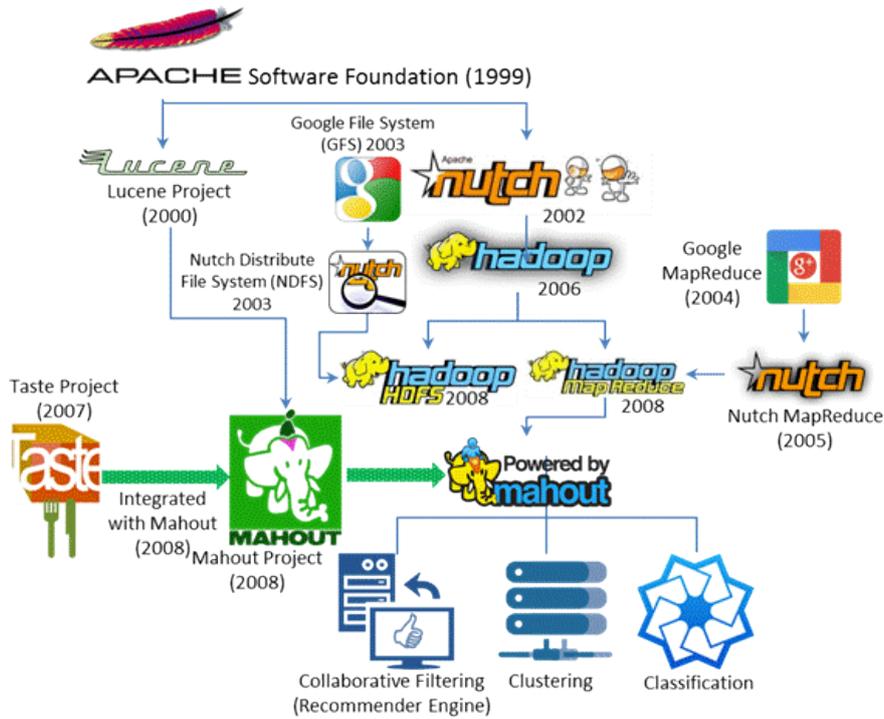

Figure 14: Evolution of GFS, HDFS MapReduce and Hadoop

### 1.7.3     The Origin of Hadoop Project

*Lucene*

According to Otis Gospodnetic et al [63], "Lucene is a high performance scalable Information Retrieval (IR) library. It lets developers add indexing and searching capabilities to their applications. Lucene was a mature, free, open-source project implemented in Java. It's a member of the popular Apache Jakarta family of projects, licensed under the liberal Apache Software License" (see Figure 15). It was written by Doug Cutting in 2000 in Java. In Sep 2001, Lucene was absorbed by Apache Software Foundation (ASF).

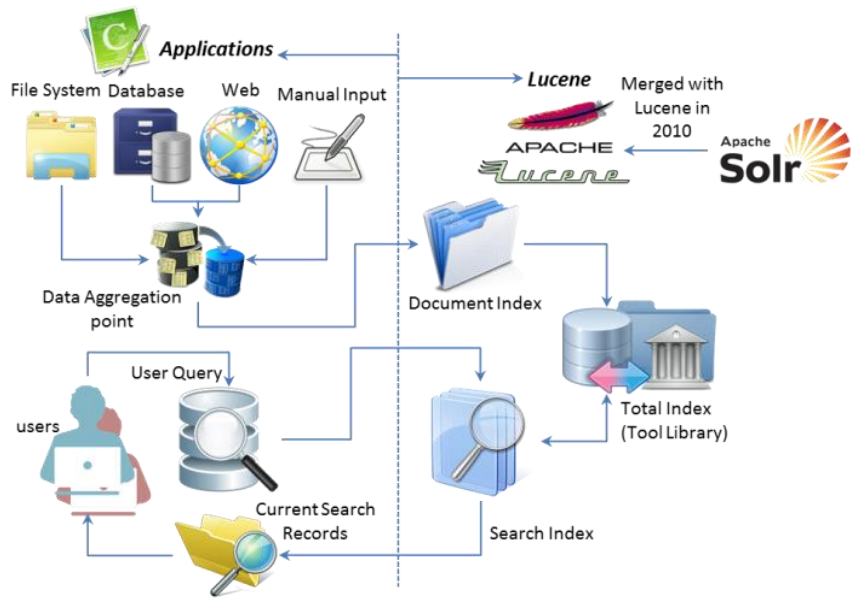

Figure 15: Connection between Apache Lucene and other Applications

However, Lucene was not an executable application or search engine rather than a toolbox or searching tool kits that enable many applications to borrow or use it. Lucene is just classification index. It converts any data to a textual format and enables them to be searchable. Its powerful searching capability is beneficial many third parties. At the heart of Lucene IR library, it is the searching and indexing capability. In order to utilize Lucene's searching and indexing functions, another open source software - Nutch is required, which was also built by Doug Cutting in 2002 (see Figure 14).

*Nutch*

Nutch is the predecessor of Hadoop, which is an open-source and executable search engine file system. There are two main reasons to develop Nutch:

- Create a Lucene index (web crawler)
- Assist developers to make query of their index

There are a lot of codes in Nutch program (such as HTTP fetcher and URL database). Michael J. Cafarella indicated that the text searching was the center piece of any search engine or web crawler, which was included in Nutch.

Based on Zakir Laliwala et al [64], another Apache project, namely Solr was developing the similar searching function as Nutch in parallel. It was also an open source enterprise platform for full text search, which was initiated by CNET in 2004. It became an Apache project in 2007. Since then, it has absorbed many tools in Apache Lucene's library to enhance and extend its full text search capability. Like Apache Lucene, it was not an executable search application rather than a toolkit or information retrieval (IR) library [65]. Therefore, Solr and Lucene had been merged into a single development project since 2010 [66].  As shown in Figure 15, although both Lucene and Solr had adopted many techniques for index searching, text mining and information retrieval algorithms, many techniques can be generalized as classification algorithms.

In general, BDA applications need different algorithms or techniques, such as clustering, collaborative filtering (or recommender engine) and others. These requirements lead to the beginning of Mahout Project in 2008 as a subproject of Apache Lucene. Since all the algorithms of both Lucene and Mahout are closely associated with the concept of machine learning, In Apr-2010, Mahout has risen as a top level project in its own right.

*Mahout*

The original object of Mahout was to build a Java-based machine learning library that covers all machine learning algorithms or techniques in theory but it can mainly handle three types of machine learning algorithms in practice:

- Collaborative filtering (Recommender Engines)
- Clustering and
- Classification

If other learning algorithms are required, we have to check the Apache Mahout URL [67] and to find out whether MapReduce can support particular algorithm or not before this algorithm can be applied in a large scalable environment. In other words, Mahout is not a universal ML library. In addition of scalable issue, Hadoop is very slow for ML workloads.  It led to the development of complimentary ecosystems, such as Hama, Storm, Spark, and Flink that addressed weakness of MapReduce-based systems.

### 1.7.4    Spark and Spark Stack

Spark was developed by UC Berkeley RAD Lab (now called as AMP Lab).  The main contributor is Matei Zaharia et al [68] [69]. ]. Its original objective was to extend Hadoop to a general purpose framework that adopts Resilient Distributed Datasets (RDDs) in memory computation (micro-batch) technique. In a simple term, it intends to replace MapReduce model with a better solution. It emphasizes the computational efficiency of iterative and recursive algorithms and interactive queries of data mining. It claimed that it would be 10-20X faster than MapReduce for certain type of workload, such as performing iterative algorithm.

Although it attempts to replace MapReduce, it did not abandon HDFS. It leverages Hadoop's file storage system. Like many other Hadoop related projects, it is an open source project under Apache Software Foundation (ASF). In June 2013, it was moved to ASF as an incubator. Since 2014, it has become an Apache top level project and supported by many Big Data vendors, such as Cloudera, Horton, SAP and MapR as noted in Figure 16.

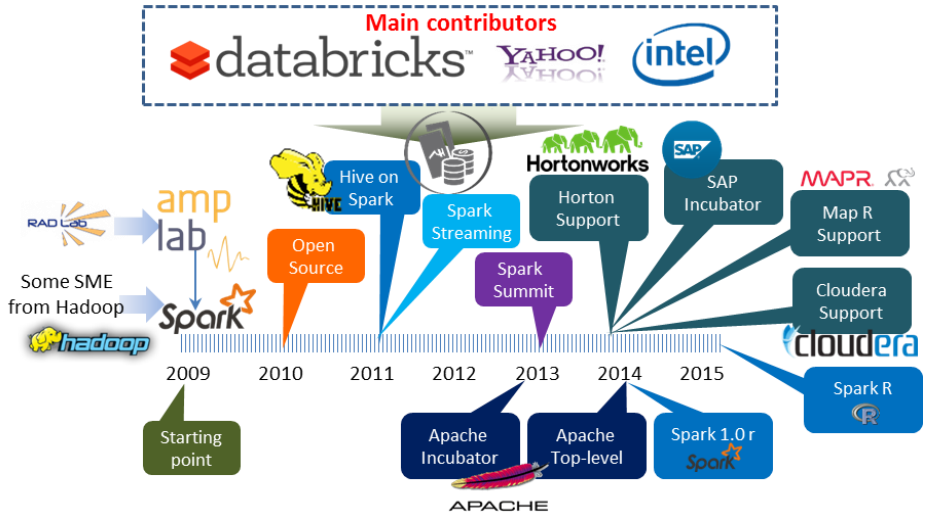

Figure 16: Spark History

Generally, Spark is a fast and general- purpose computation platform based on large clusters. In contrast to MapReduce that is basically designed for web crawler, indexing system and limited machine learning, Spark includes SQL, interactive query, data stream, graph, and machine learning analytic functions into its computation platform.

Based on the Berkeley Data Analytics Stack (BDAS) architecture, Spark developed as a unified stack integrating all libraries and higher level components together (see Figure 17). Spark consists of seven major elements: Spark core of data engine, Spark cluster manager (includes Hadoop, Apache Mesos and built-in Standalone cluster manger), Spark SQL, Spark streaming, Spark Machine Learning Library, Spark GraphX, and Spark programming tools.

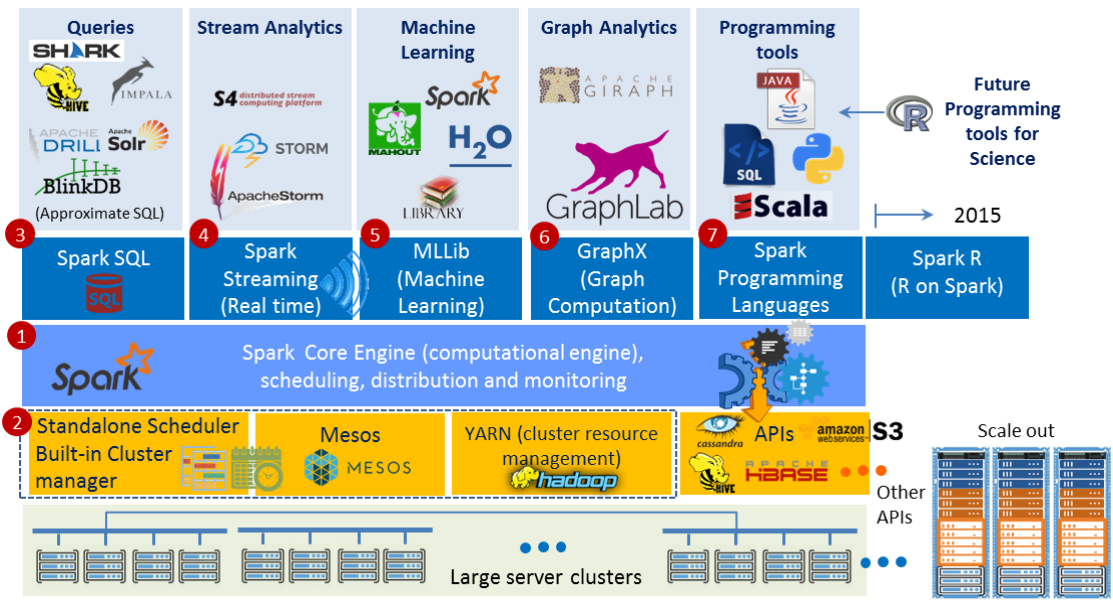

Figure 17: SPARK Analytic Stack

### 1.7.5 Flink and other data process engines

Apart from Spark, there are several data processing engines such as Microsoft Dryad, Storm, Tez, Flink and CIEL (see Figure 18) that are capable of supporting MapReduce like processing requirements. They aim to support more computational functions, such as standard queries, stream analysis, machine learning, graphic analysis and interactive or ad hoc queries efficiently. The effort made by these platforms is to generalize Hadoop to be able to support a wide variety of BDA workloads.

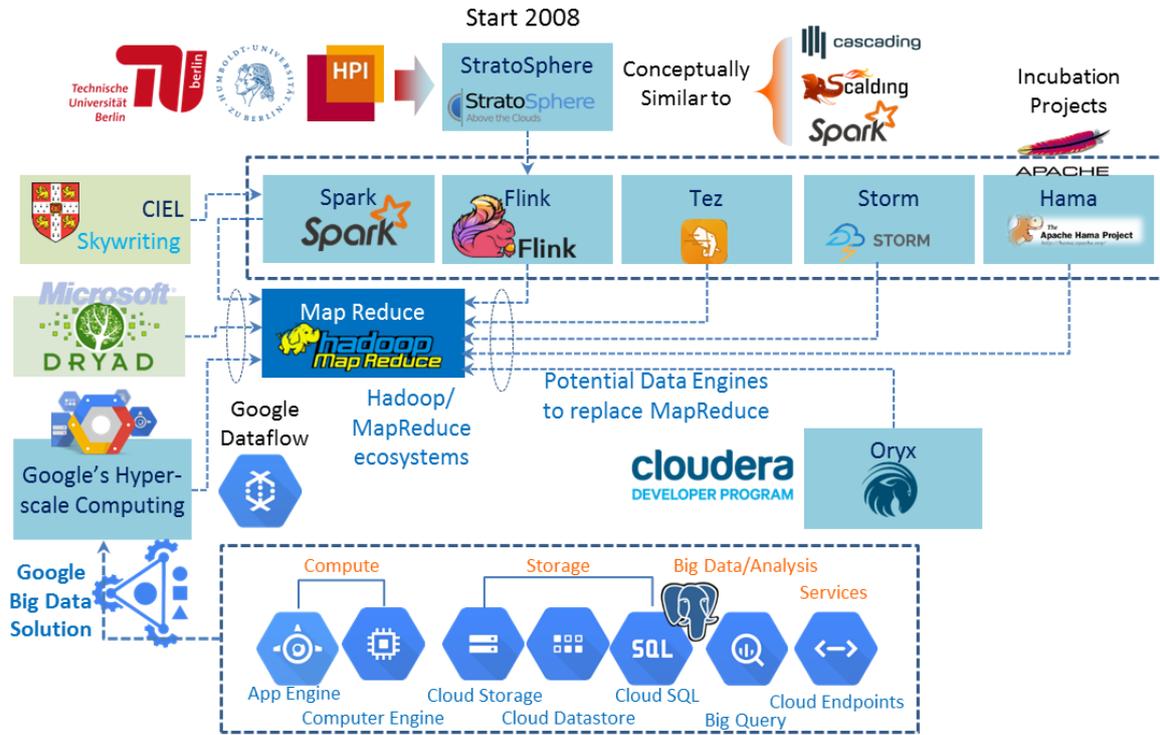

Figure 18: Potential data processing engines to replace MapReduce

Stephan Ewen et. al. [70], Kostas Tzoumas [71] and Marton Balassi [72] argued that Flink is the next generation or the 4th generation data processing engine in comparison with others (see Table 6 and Figure 19) although each data processing engine has its own special feature. Flink data engine is truly general purpose framework for Big Data Analytics (BDA). They claim that Flink is capable of outperforming Spark by 2.5 times.

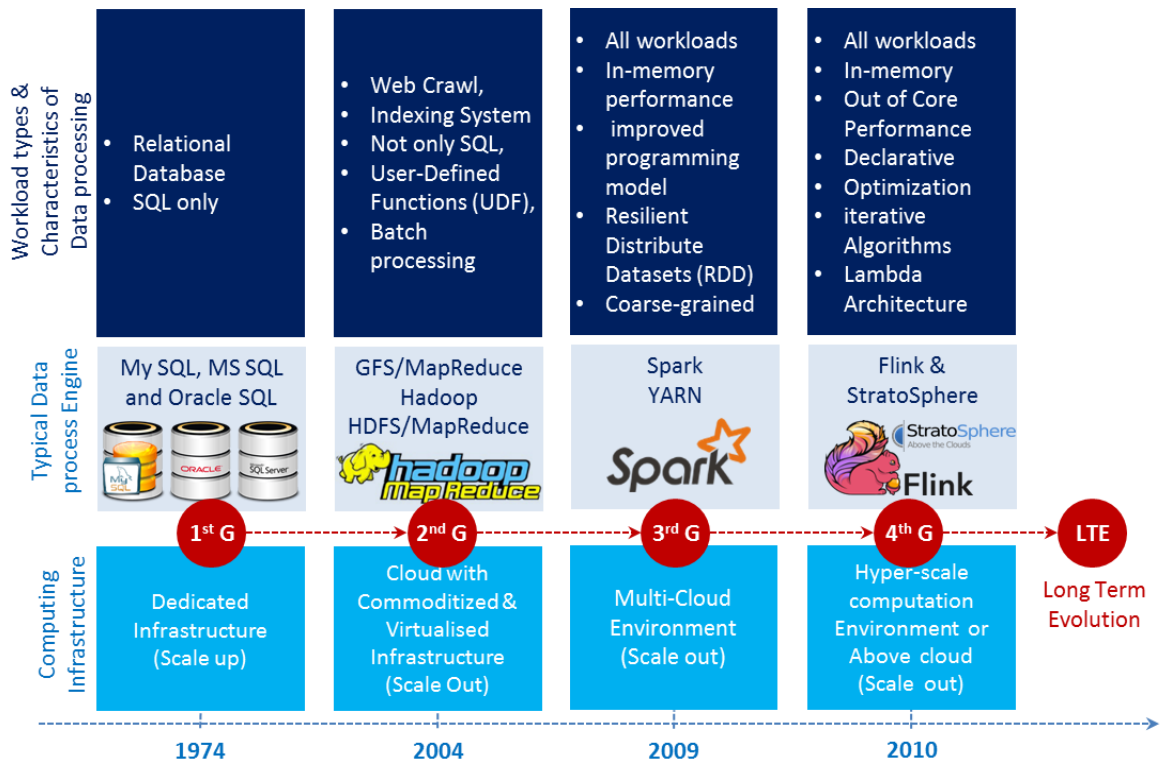

Figure 19: Evolution of Data and Big Data process engines

| Data process engines comparison | MapReduce 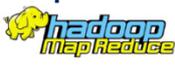 | Tez 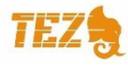 | Spark 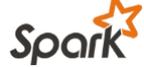 | Flink 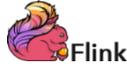 |
|---|---|---|---|---|
| Start at | 2004 | 2007 | 2009 | 2010 |
| API | MapReduce on Key/Value pairs | Key/Value pair Readers/Writers | Transformations on key/value pair collections | Iterative transformations on collection or iteration aware |
| Paradigm | MapReduce | Direct Acyclic Graph (DAG) | Resilient Distributed Datasets (RDD) | Cyclic data flows or dataflow with feedback edges |
| Optimization | none | none | Optimization of SQL queries | Optimization in all APIs |
| Execution | Batch | Batch sorting and partitioning | Batch with memory pinning | Stream with out of core algorithms |
| Enhanced features plus Specialise particular workloads | • Small recoverable tasks, • Sequential code inside map & reduce functions | • Extends map/reduce model to DAG model • Backtracking-based recovery | • Functional implementation of Dryad recovery (RDDs) • Restrict to coarse-grained transformations • Direct execution of API | • Embed query processing runtime in DAG engine • Extend DAG model to cyclic graphs • Incremental construction of graphs |

Table 6: Data Processing Engine Comparison

A possible reason for Ewen et. al. to claim that Flink is better than Spark is that it is based on Lambda architecture and able to process arbitrary Big Data workloads in real time. The basic concept of Lambda architecture is to build the data processing engine or system with the number of layers in order to deal with a subset of data with stream properties. These layers are only few thousand line of code to implement a total of seven steps (2 for batch layer, 2 for serving layer and 3 for speed layers, see Figures 20 and 21).

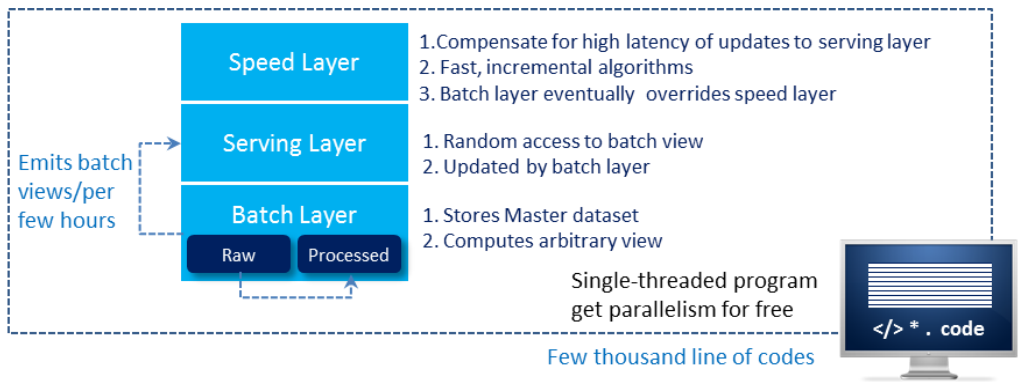

Figure 20: The Process steps of Lambda Architecture

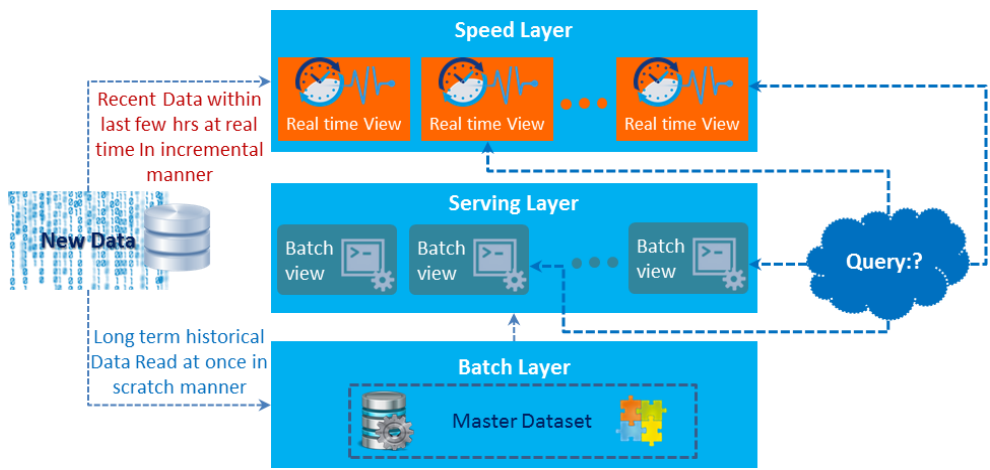

Figure 21: The Elements of Lambda Architecture

The purpose for establishing these three layers, according to Nathan Marz [73], is to meet the characteristic requirements of all types of Big Data workloads. They are:

- Robustness and fault tolerance
- Low latency reads and updates
- Scalability
- Generalization
- Extensibility
- Ad hoc queries
- Minimal maintenance
- Debuggability

Figure 22 shows that the batch layer as a part of Hadoop can easily meet robustness and fault tolerance requirements. Scalability is the requirement for both batch and serving layers that both Hadoop and Elephant DB can handle it. Extensibility means data stream adds a new function to the master dataset. The batch layer allows users to recompute another entire batch view from scratch. To some extent, this also means that batch layer can perform ad hoc queries because the master dataset in one location. Due to the nature of Hadoop's robustness, minimal maintenance is acceptable because a serving layer database only gets a batch view per few hours, which emitted from batch layer. In other words, it doesn't write randomly very often and has so few moving parts. Subsequently, it is less likely to go wrong.

The combination of both batch and serving layers can record all intermediate steps of outputs (serving layer) and inputs (batch layer – master dataset) for data process. Therefore, if the process has any hiccup, the debug analysis is quite easier.

The top element of the Lambda architecture is the speed layer. The purpose of having speed layer is to perform arbitrary computing function on arbitrary data in real time, which is to fill the gap time of new data for both batch and serving layers that have been left with. In contrast to batch layer, the speed layer only checks the latest data while batch layer covers all the data in one batch.  Moreover, it only does in an incremental manner rather than in a re-compute from scratch manner that the batch layer does. The speed layer capability meets the Big Data requirements for low latency reads and updates.

The overall Big Data query is the combination of real time and batch views as noted in Figure 22, which shows an example query processing system based on Lambda architecture.   In contrast to MapReduce (batch only), the Lambda architecture can meet all requirements of Big Data query whether it is batch or real time.

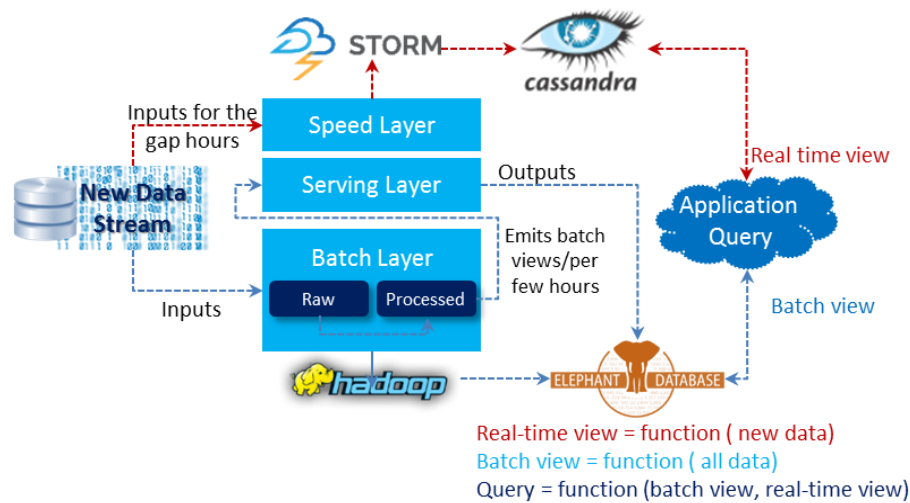

Figure 22: An Example of Implementation of Lambda Architecture

In addition to Flink and Spark, more than 40 processing engines are available, which are capable of processing different types of BDA workloads (see Table 7).

| Bashreduce | Gearman | Minceat | R3 |
|---|---|---|---|
| Ceph | GoCircuit | Mincemeat | Riak |
| Cloud MapReduce | GPMR | Misco | SAMOA |
| Cloud-Crowd | HaLoop | MongoDB | Skynet |
| Condor | HPCC | Octopy | Spark |
| Data Core | HTTPMR | Oryx | Sphere |
| DISCO | Aneka MapReduce | Plasma MapReduce | Storm |
| Elastic Phoenix | MapRedus | Preregrine | Tachyon |
| Filemap | MapRejuice | QFS | TEZ |
| Flink (Stratosphere) | MARS | Qizmt | Weka |

Table 7: 40 Alternative Platforms for Big Data Processing.

### 1.7.6 Summary of Hadoop and its Ecosystems

Hadoop has become the standard framework to run distributed BDA that can process massive scale of data on large clusters based on the commodity hardware or a cloud infrastructure. Along with its evolutionary journey, it has absorbed and integrated some Apache projects that have similar functionalities, such as Taste, Solr and Mahout. Due to the demand for processing all types of BDA workloads, many Hadoop's ecosystems have been developed, such as Spark, Storm, Hama, Tachyon, TEZ, S4 and Flink. These ecosystems intend to overcome MapReduce's shortcomings and to specialize with particular type of BDA workload. Consequently, some platforms have been generalized to handle all types of BDA workloads.

*Hadoop Key Functions*

When Douglas Cutting and Michael J. Cafarella created it in early 2006, their original idea was to build Apache Nutch (or a web crawler engine) on a cheaper infrastructure. It consists of five key functional components (see Figure 5):

1. ETL tools for data integration
2. Functional element or programming model: MapReduce
3. Core units: Distributed Framework or storage system
4. Processing modules or libraries: Machine Learning
5. Administration models

In comparison with many other conventional databases, Hadoop is not a database but a distributed storage and computational framework. It is a free and open source ecosystem. It has six characteristics:

1. Scale-out with distributed computing
2. Expect failure with redundancies
3. Smart software with dumb hardware
4. Share nothing architecture
5. Move processors not data (taking computer to data, rather than other way around)
6. Building applications, not infrastructure

*Hadoop's Distinguish Features*

One of the Hadoop's unique features is that it is supported by so many auxiliary tools, especially for many administration tools, such as monitoring, management and maintenance (see Figure 9). It also has many APIs to interface with other BDA applications. Many ASF incubation projects (such as Spark and Flink) can replace MapReduce but it would be too costly to substitute the entire Hadoop framework.

# 1.8 ML + CC → BDA and Guidelines

We discussed the role of machine learning (ML), Cloud Computing (CC), and Hadoop like systems. We see that ML and CC are the two most important components of BDA. If there are no advances in ML and CC, BDA could not be implemented or operated cost effectively. Of course BDA needs good understanding of application domain. Effective BDA needs appropriate choice of ML techniques and the use of CC to handle big data sets for both training and extracting new meaningful data patterns. CC can provide an affordable solution for many individuals and small to medium scale enterprises. Therefore, we assert that ML + CC → BDA. Hadoop's history and its ecosystems with machine learning applications have demonstrated this concept adequately.

Finally, BDA is not an ideal solution for every analytics problem. For some cases, it may only add the burden to the business. Table 8 provides guidelines to decide which cases could be applied for BDA solutions and which ones would not benefit from BDA. These guidelines help in determining the case for BDA.

| 3 Aspects | 9 Vs | Fit for Big Data Analytics | Not Fit for Big Data Analytics |
|---|---|---|---|
| **Data** | Volume | • Datasets do not fit into one node (e.g. PB –EB size datasets)<br>• Bringing computing to the data | • Dataset can be fit into one node<br>• Bringing data to the computing node |
| | Variety | • Not only SQL<br>• Collection and discovery of datasets from different data sources (e.g. M2M, WSN, RFID, SCADA, SQL and NoSQL)<br>• Schema [3] on read | • One type workload (RDBMS or SQL)<br>• Single data source<br>• Schema on write |
| | Velocity | • Data agility<br>• Interactive and dynamic data stream | • Traditional stable environment<br>• Static dataset |
| **Statistics** | Veracity | • Datasets are not clean<br>• Models construction need many "What-if" for fidelity issues<br>• Rely on archived data for reliability and credibility | • Dataset is relatively clean<br>• Model construction is relatively simple<br>• Require live data |
| | Variability | • Heterogeneous Dataset<br>• Dynamic or flexible schemas<br>• Numerous variables of dataset (e.g. > 50 variables) | • Homogeneous Dataset<br>• Fixed schema<br>• Few variables of dataset |
| | Validity | • Require independent and transparent criteria to verify the result of BDA (e.g. GFT) | • Simple and straightforward approach to verify the result of data mining |
| **Business Intelligence** | Value | • Solving strategic problems that have long term consequences (e.g. competitive advantages, integrity, excellence, sustainability, success and cost leadership)<br>• Leveraging business values from all data sources | • Routine issues for a short term<br>• Exploring business value from single source |
| | Verdict | • Ask for an answer | • Ask for the answer |
| | Visibility | • Search for strategic insight | • Search for temporary solutions |
| **Other Aspects** | | • Large scale computing needing high fault tolerance<br>• Scale out<br>• High percentage of parallel and distributed processing workload | • Fault tolerance may not be essential<br>• Scaling up<br>• Percentage of serial processing workloads is higher |

Table 8: Guidelines for Big Data Analytics

# 1.9 Conclusion

We have highlighted many major events and debates in Big Data and introduced the original concept of Big Data and its 3Vs attributes. We proposed an extension to this view of Big Data from 3Vs to $3^2$Vs (9 Vs) to capture the full meaning of BDA to include additional attributes of Business Intelligence and Statistics aspects (see Figure 23). We provided an overview of many popular platforms for BDA such as Hadoop, Spark and Flink that are affordable to small and medium scale enterprises. We have developed the notion that ML + CC → BDA. That is, the execution of machine learning tasks on large-data sets in cloud computing environments is often called as Big Data analytics.

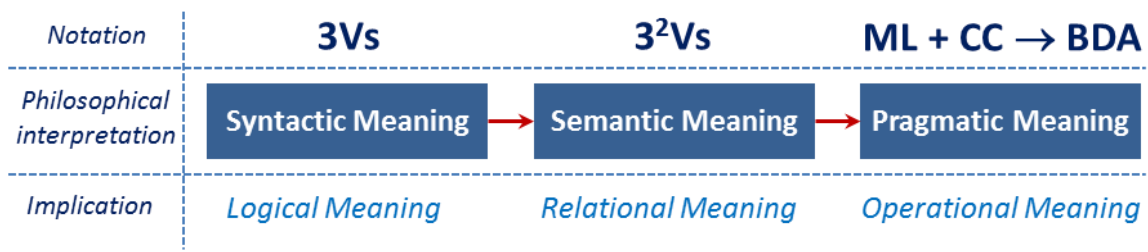

Figure 23: Comprehensive meaning of Big Data

---

[3] "Schema-on-Read" means a table or a set of statements is not pre-defined. Sometime it is also named as "Schemaless" or "Schema free" In contrast, "Shema-on-Write" means that a table is predetermined. Sometime, it is also called as "fixed schema"[77] [78] [79]